\documentclass[11pt]{article}
\usepackage{amssymb}
\usepackage{amsmath}
\usepackage{amscd}
\usepackage{latexsym}

\catcode`@=11
\renewcommand{\section}{\@startsection{section}{1}{0pt}{\medskipamount}
{\medskipamount}{\large\bf}}
\numberwithin{equation}{section}
\catcode`@=12
\def\a{\alpha}
\def\b{\beta}
\def\g{\gamma}

\def\eps{\epsilon}
\def\ve{\varepsilon}
\def\z{\zeta}
\def\h{\eta}
\def\th{\theta}

\def\m{\mu}

\def\j{\psi}

\def\La{\Lambda}

\def\tx{\tilde x}
\def\ty{\tilde y}
\def\tt{{\tilde\theta}}

\def\1{\dot 1}
\def\2{\dot 2}

\def\eps{\epsilon}

\newcommand{\ab}{{\bar{\a}}}
\newcommand{\bb}{{\bar{b}}}

\newcommand{\wb}{{\bar{w}}}
\newcommand{\zb}{{\bar{z}}}

\newcommand{\C}{\mathbb C}
\newcommand{\R}{\mathbb R}

\newcommand{\Hcal}{{\cal H}}
\newcommand{\Acal}{{\cal A}}
\newcommand{\Bcal}{{\cal B}}
\newcommand{\Ccal}{{\cal C}}
\newcommand{\Ncal}{{\cal N}}

\newcommand{\Fcal}{{\cal F}}

\def\im{\textrm{i}}
\def\ep{\textrm{e}}
\def\N2{$N{=}2$}
\def\pa{\mbox{$\partial$}}

\def\sfrac#1#2{{\textstyle\frac{#1}{#2}}}
\def\>{\rangle}
\def\<{\langle}
\def\+{\dagger}
\def\={\ =\ }
\def\und{\qquad\textrm{and}\qquad}

\begin{document}

\begin{titlepage}
\setcounter{page}{0}
\begin{flushright}
ITP--UH--21/07\\
\end{flushright}

\vskip 2.0cm

\begin{center}

{\Large\bf  Scattering of Noncommutative Waves and Solitons in a \\[5mm]
            Supersymmetric Chiral Model in 2+1 Dimensions
}

\vspace{15mm}

{\large Christian Gutschwager${}^*$,  Tatiana A. Ivanova${}^\+$ \ and \ 
Olaf Lechtenfeld${}^*$ }
\\[10mm]
\noindent ${}^*${\em Institut f\"ur Theoretische Physik,
Leibniz Universit\"at Hannover \\
Appelstra\ss{}e 2, 30167 Hannover, Germany }\\
{Email: gutschwager@math.uni-hannover.de, lechtenf@itp.uni-hannover.de}
\\[10mm]
\noindent ${}^\+${\em Bogoliubov Laboratory of Theoretical Physics, JINR\\
141980 Dubna, Moscow Region, Russia}\\
{Email: ita@theor.jinr.ru}
\vspace{15mm}

\begin{abstract}
\noindent
Interactions of noncommutative waves and solitons in 2+1 dimensions can be 
analyzed exactly for a supersymmetric and integrable U($n$) chiral model 
extending the Ward model. Using the Moyal-deformed dressing method in an
antichiral superspace, we construct explicit time-dependent solutions of 
its noncommutative field equations by iteratively solving linear equations. 
The approach is illustrated by presenting scattering configurations for 
two noncommutative U(2) plane waves and for two noncommutative U(2) solitons 
as well as by producing a noncommutative U(1) two-soliton bound state.
\end{abstract}

\end{center}
\end{titlepage}

\section{Introduction and summary}
\noindent
Solitonic solutions of the field equations of motion play an essential role in 
our understanding of field and string theories beyond perturbation theory. 
This persists for the noncommutative extension of scalar and 
supersymmetric gauge field theories which appear naturally 
in string theories~\cite{DH}--\cite{SW}.
In particular, the massless modes of the open $N$=2 string in a space-time 
filling brane with a constant NS $B$-field are described by noncommutative 
self-dual Yang-Mills (SDYM) theory in 2+2 dimensions~\cite{LPS1}. 
Upon reduction on the worldvolume of $n$ coincident D2-branes there emerges
a noncommutative generalization~\cite{LPS2} of a modified U($n$) chiral model 
in 2+1 dimensions known as the Ward model~\cite{Ward88}.
The integrability of this model~\cite{Ward88}--\cite{TU} is preserved
under the noncommutative deformation~\cite{LP1, LP2}. 
In~\cite{LP1}--\cite{Klawunn} families of multi-soliton and plane-wave 
solutions to its Moyal-deformed equations of motions were studied. 
Also, reductions of wave configurations in the 2+1 dimensional
model to solutions of the noncommutative sine-Gordon equations were 
described~\cite{Penati}. We remark that not only the (noncommutative) 
Ward-model or sine-Gordon equations but a lot of other integrable equations 
in three and fewer dimensions derive from the (noncommutative) SDYM equations 
by suitable reductions
(see e.g.~\cite{Ward}--\cite{Dimakis} and references therein).

Given the fact that spacetime supersymmetry is an essential ingredient of
string theory, it is natural to consider supersymmetric extensions of the
above scenario. This was done by Witten~\cite{Wit} who has shown that 
$\Ncal{=}4$ super SDYM theory appears in twistor string theory.\footnote{
For further developments and references see e.g.~\cite{Popov, Bedford}.}
Later, it was shown that $\Ncal{=}4$ super SDYM in
2+2 dimensions can be reduced to an $\Ncal{=}8$ supersymmetric extension
of the Ward model in 2+1 dimensions~\cite{sigma8}. Subsequently,
truncations to $\Ncal{<}8$ and a noncommutative Moyal deformation of 
this model were considered, and noncommutative multi-soliton solutions were 
constructed~\cite{LP07}. However, as in the non-supersymmetric case, generic 
multi-soliton configurations were found to be devoid of scattering~\cite{LP07}
(see also~\cite{Ohta}).

These supersymmetric no-scattering soliton configurations were obtained by 
applying a solution-generating technique (the dressing method~\cite{Z}) to the 
$\Ncal$-extended noncommutative U($n$) Ward model, taking a dressing ansatz for
the $\j$ function with only first-order poles in the spectral parameter $\z$. 
Here we show that for multi-{\it waves\/} this ansatz yields nontrivial
wave-wave interactions since each plane wave experiences a phase shift. 
Furthermore, by allowing for {\it second-order\/} poles in the dressing ansatz,
we construct $\Ncal$-extended noncommutative (time-dependent) two-soliton 
configurations with genuine soliton-soliton interaction.
Thus, the studied features of the undeformed Ward model survive not only
the Moyal deformation but the supersymmetric extension as well.

\bigskip
   
\section{The noncommutative $\Ncal$-extended Ward model}
\noindent
Recall that nonlinear sigma models in 2+1 dimensions\footnote{
Sigma models in $k$ dimensions describe mappings of a $k$-dimensional 
manifold~$X$ into a manifold~$Y$. Chiral models pertain to the special case 
when $Y$ is a Lie group.} 
may be Lorentz-invariant or integrable but not both. 
An integrable model appears when one adds to the standard sigma-model 
field equations a Wess-Zumino-Witten term which explicitly breaks the 
Lorentz group SO(2,1) to the group SO(1,1)~$\cong$~GL(1, $\R$)~\cite{Ward88}. 
An $\Ncal{=}8$ supersymmetric generalization of this model has been 
introduced in~\cite{sigma8} and is easily truncated to any smaller even 
number~$\Ncal$ of supersymmetries. 
To formulate this model, one should introduce: %\\[-22pt]
\begin{itemize}
\addtolength{\itemsep}{-6pt}
\item the space $\R^{2,1}=(\R^3, g)$ with coordinates $(x^a)=(t,x,y)$ 
\item the metric $g=\textrm{diag}(-1, +1, +1)$ 
\item the superspace $\R^{3|2\Ncal}$ with coordinates $(x^a|\h^\a_i,\th^{i\a})$
for $\a = 1,2$ and $i=1,\ldots,\sfrac12\Ncal\le 4$
\item the antichiral superspace $\R^{3|\Ncal}$ with coordinates $(x^a|\h^\a_i)$
\end{itemize}
The $\Ncal$-extended Ward model describes the dynamics of a U($n$)-valued 
superfield $\Phi (x^a, \h^\a_i)$ living on the antichiral 
superspace~$\R^{3|\Ncal}$. The noncommutative Moyal deformation of this model 
was considered in~\cite{LP07}. 

Since integrability can be preserved in noncommutative deformations (see 
e.g.~\cite{LP1}--\cite{Penati}), we right away Moyal deform the supersymmetric
Ward model with a constant real noncommutativity parameter $\th\ge0$. 
This is achieved by replacing the ordinary product of classical fields (or 
their components) with the noncommutative associative star product,\footnote{
See~\cite{reviews} for reviews on noncommutative field theories.}
\begin{equation}\label{2.1}
(f\star g)(t,x,y,\h_i^\a) \= f(t,x,y,\h_i^\a)\,\exp \{\sfrac{\im}{2}\th\,
(\overleftarrow{\pa_x}\overrightarrow{\pa_y}-
 \overleftarrow{\pa_y}\overrightarrow{\pa_x})\}\,g(t,x,y,\h_i^\a)\ .
\end{equation}
Note that we choose a purely bosonic space-space deformation,
i.e. the time coordinate remains commutative and no derivatives with respect 
to the Grassmann variables $\h_i^\a$ appear in~(\ref{2.1}).
The U($n$)-valued superfield $\Phi(t,x,y,\h^\a_i)$ of the noncommutative
$\Ncal$-extended U($n$) Ward model~\cite{LP07} obeys the classical field 
equations
\begin{equation}\label{2.2}
\begin{aligned}
&\pa_x(\Phi^\+\star\pa_x\Phi) + \pa_y(\Phi^\+\star\pa_y\Phi)
- \pa_t(\Phi^\+\star\pa_t\Phi) + \pa_y(\Phi^\+\star\pa_t\Phi)
- \pa_t(\Phi^\+\star\pa_y\Phi) \=0\ ,\\[2pt]
&\pa_1^i(\Phi^\+\star\pa_x\Phi) - \pa_t(\Phi^\+\star\pa_2^i\Phi)
+ \pa_y(\Phi^\+\star\pa_2^i\Phi) \=0\ ,\\[2pt]
&\pa_1^i(\Phi^\+\star\pa_t\Phi) + \pa_1^i(\Phi^\+\star\pa_y\Phi)
- \pa_x(\Phi^\+\star\pa_2^i\Phi) \=0\ ,\\[2pt]
&\pa_1^i(\Phi^\+\star\pa_2^j\Phi) + \pa_1^j(\Phi^\+\star\pa_2^i\Phi) \=0\ ,
\end{aligned}
\end{equation}
where $\pa^i_\a:=\pa /\pa\h_i^\a$, and the unitarity condition reads
$\Phi^\+\star\Phi=\Phi\star\Phi^\+={\bf 1}_n$, 
with $\+$ denoting hermitian conjugation. 
The Wess-Zumino-Witten terms responsible for the integrability 
are the last two terms in the first line above.

As it was discussed in~\cite{sigma8, LP07}, the field equations (\ref{2.2})
are equivalent to 
\begin{equation}\label{2.5}
\hat D^i_{\a}\hat\Acal_{\b}^j + \hat D^j_{\b}\hat\Acal_{\a}^i +
\hat \Acal^i_{\a}\star\hat\Acal_{\b}^j+\hat A^j_{\b}\star\hat\Acal_{\a}^i
+ (\a \leftrightarrow\b ) \= 0
\end{equation}
with the full superfields
\begin{equation}\label{2.6}
\hat\Acal^i_{1}\=0  \und \hat\Acal^i_{2}\= \Phi^\+\star\hat D^i_{2}\Phi \ .
\end{equation}
Here,
\begin{equation}\label{2.7}
\hat D_{\a}^i =\pa_{\a}^i+2\th^{i\b}\pa_{(\a\b )}\quad\mbox{with}\quad
\pa_{(11)}=\pa_{t}-\pa_y\ ,\
\pa_{(12)}=\pa_{(21)}=\pa_x\quad\mbox{and}\quad
\pa_{(22)}=\pa_{t}+\pa_y\  .
\end{equation}
Furthermore, by expanding the superfields $\hat\Acal^i_\a$ in $\h_j^\b$, 
one can show that the equations~(\ref{2.5}) are equivalent to a supersymmetric
extension of the Bogomolny-type Yang-Mills-Higgs equations on a multiplet 
$(A_{(\a\b)}$, $\chi^{i\a}$, $H$, $\phi^{ij}$, $\tilde\chi_i^\a$, $G_{\a\b})$ 
of space-time component fields or its $\Ncal{<}8$ truncation
\cite{sigma8, LP07}. Here 
$A_{(\a\b)}=A_{(\b\a)}$ are the components of a vector in spinorial notation,
$G_{\a\b}=G_{\b\a}$ are the components of a (pseudo)vector dual to a two-form, 
$\chi^{i\a}$ and $\tilde\chi_i^\a$ are components of spinors, 
and $\phi^{ij}=-\phi^{ji}$ are scalars in addition to the Higgs scalar~$H$.

The supersymmetry algebra in 2+1 dimensions is generated by the 
$2\Ncal$ supercharges
\begin{equation}\label{2.9}
\hat Q_{i\a}=\pa_{i\a}-2\h_i^{\b}\pa_{(\a\b)} \und
\hat Q^i_{\a}=\pa^i_{\a}\ .
\end{equation}
The extended and deformed Ward-model equations (\ref{2.2}) are invariant under
the infinitesimal supersymmetry transformations generated by these supercharges 
because they anticommute with~$\hat D_{\a}^i$. 

In order to avoid cluttering the formulae we suppress the `$\star$' notation 
for noncommutative multiplication from now on; all products are assumed to be 
star products, and functional operations (e.g. inverses) use the star product.

\bigskip

\section{Explicit solutions via the dressing approach}
\noindent
{\bf Linear system.} \ 
One of the powerful tools for constructing solutions to integrable 
equations is the so-called `dressing method'~\cite{Z} which is easily extended 
to the noncommutative and supersymmetric setup~\cite{LP1, LP2, LP07}. 
The key observation is that the field equations (\ref{2.5}) (or (\ref{2.2})) 
can be obtained as compatibility conditions for a linear system of 
differential equations. The six antichiral superfield components
\begin{equation}\label{3.1}
\hat\Acal^i_{\a}\= \Acal^i_{\a} + 2\th^{i\b}(\Acal_{(\a\b)}-\ve_{\a\b}\Hcal)
\end{equation}
defined by (\ref{2.6}) and (\ref{2.7}) read
\begin{equation}\label{3.2}
\begin{aligned}
\Acal_{(12)}-\Hcal&=&0 \und\qquad
\Acal_{(22)}\=\Phi^\+\pa_{(22)}\Phi \ =:\ \Acal \\[2pt]
\Acal_{(11)}&=&0 \und
\Acal_{(12)}+\Hcal\=\Phi^\+\pa_{(12)}\Phi \ =:\ \Bcal \\[2pt]
\Acal^i_1&=&0 \und\qquad\quad
\Acal^i_2\=\Phi^\+\pa^i_2\Phi \quad\ =:\ \Ccal^i
\end{aligned}
\end{equation}

With these data, we consider the linear equations
\begin{equation}\label{3.4}
\begin{aligned}
(\zeta\pa_x - \pa_t - \pa_y)\j &\= \Acal\;\j\ ,\\[3pt]
(\zeta\pa_t - \zeta\pa_y - \pa_x )\j &\= \Bcal\;\j\ ,\\[2pt]
(\zeta\pa_{1}^i - \pa_{2}^i)\j &\= \Ccal^i\,\j\ ,
\end{aligned}
\end{equation}
where the $n{\times}n$ matrix $\j$ depends on $(x^a|\zeta,\eta_i^\a)$ 
and the $n{\times}n$ matrices $\Acal$, $\Bcal$ and $\Ccal^i$ are superfield 
functions of $(x^a|\eta_i^\a)\in \R^{3|\Ncal}$ but do not depend on the 
spectral parameter~$\zeta$
which lies in the extended complex plane $\C\cup\{\infty\}=\C P^1$.

\medskip

\noindent
{\bf Compatibility conditions.} \ 
The compatibility conditions for the linear system (\ref{3.4}) are
\begin{equation}\label{3.7}
\begin{aligned}
&\pa_x\Acal -(\pa_t{+}\pa_y)\Bcal - [\Acal,\Bcal ]\=0\ , \quad 
\pa_2^i\Acal - (\pa_t{+}\pa_y)\Ccal^i + [\Ccal^i,\Acal]\=0\ ,\\[2pt]
&\pa_2^i\Bcal -\pa_x \Ccal^i + [\Ccal^i,\Bcal]\=0\ ,\quad
\{\pa_2^i+\Ccal^i, \pa_2^j+\Ccal^j\}\=0\ ,\\[2pt]
&(\pa_t{-}\pa_y)\Acal - \pa_x\Bcal \= 0\ ,\quad 
\pa_1^i\Bcal - (\pa_t{-}\pa_y)\Ccal^i \=0\ ,\\[2pt]
&\pa_1^i\Acal - \pa_x\Ccal^i \=0\ ,\quad 
\pa_1^i\Ccal^j+\pa_1^j\Ccal^i \=0\ .
\end{aligned}
\end{equation}
It is easy to see that (\ref{3.2}) solves the first two lines
and turns the last two lines into~(\ref{2.2}).

The GL($n,\C$)-valued superfield $\j$ is subject to the 
reality condition~\cite{LP07}
\begin{equation}\label{3.11}
\j(t,x,y,\zeta,\eta) \; [\j(t,x,y,\bar\zeta,\eta )]^\+ \= {\bf 1}_n\ .
\end{equation}
Inserting the parametrization (\ref{3.2}) of 
$\Acal$, $\Bcal$ and $\Ccal^i$ into the linear system~(\ref{3.4}), 
we obtain the standard gauge-fixing conditions
\begin{equation}\label{3.12} 
\begin{aligned}
&\j(t,x,y,\eta,\zeta{\to}\infty)\={\bf 1}_n + O(\zeta^{-1})\ ,\\[2pt]
&\j(t,x,y,\eta,\zeta{\to}0)\ \=   \Phi^\+(t,x,y,\h ) + O(\zeta)\ .
\end{aligned}
\end{equation}
The second equation yields $\Phi=\j^{-1}(\zeta{=}0)$ and 
also $\Acal$, $\Bcal$ and $\Ccal^i$ via~(\ref{3.2}). 

\medskip

\noindent
{\bf Explicit $\Ncal$-extended solutions.} \ 
One can rewrite (\ref{3.4}) in the form
\begin{equation}\label{3.13}
\j (\pa_t+\pa_y - \zeta\pa_x)\j^\+ = \Acal \ ,\quad
\j (\pa_x - \zeta\pa_t + \zeta\pa_y)\j^\+ = \Bcal \quad\mbox{and}\quad 
\j (\pa_2^i - \zeta\pa_1^i)\j^\+ = \Ccal^i \ ,
\end{equation}
where the right-hand sides of (\ref{3.13}) do not depend on $\zeta$.
Therefore the left-hand sides of (\ref{3.13}) as well as the reality condition 
(\ref{3.11}) do not depend on $\zeta$, while $\psi$ is expected to be a
nontrivial meromorphic function of $\zeta$  globally defined on $\C P^1$.

We briefly recall the dressing construction. We assume that $\j$ possesses 
$m$~poles in~$\z$ at mutually distinct locations~$\mu_k$
for $k=1,\ldots,m$ in the complex lower half plane. 
One can build a solution $\j_m$ featuring $m$~{\it simple\/} poles 
at $\mu_1,\ldots,\mu_m$ by left-multiplying an ($m{-}1$)-simple-pole 
solution~$\j_{m-1}$ with a single-pole factor of the form
\begin{equation}\label{3.14}
{\bf 1}_n\ +\ \frac{\mu_m - \bar\mu_m}{\zeta - \mu_m} \ P_m(x^a,\h_i^{\a})\ . 
\end{equation}
Here, the $n{\times}n$ matrix function $P_m$ is a hermitian projector
of rank $r_m$, i.e. $P_m^\+=P_m$ and $P_m^2=P_m$, 
and therefore one can decompose
\begin{equation}\label{3.15}
P_m\=T_m\,(T_m^\+T_m)^{-1}T_m^\+\ ,
\end{equation}
where $T_m$ is an $n{\times}r_m$ matrix depending on $x^a$ and $\h_i^\a$. So, 
the iteration $\j_1\mapsto\ldots\mapsto\j_m$ yields the multiplicative ansatz 
\begin{equation}\label{3.16}
\j_m\=\prod\limits^{m-1}_{\ell=0}\Bigl({\bf 1}_n\ +\
\frac{\mu_{m-\ell}-\bar\mu_{m-\ell}}{\zeta-\mu_{m-\ell}}\ P_{m-\ell}\Bigr)
\end{equation}
which, via a partial fraction decomposition, may be rewritten in the additive 
form 
\begin{equation}\label{3.17}
\j_m\={\bf 1}_n\ +\ \sum\limits^{m}_{k=1}\frac{\La_{mk}S^\+_k}{\zeta-\mu_k}\ ,
\end{equation}
where $\La_{m k}$ and $S_k$ are some $n{\times}r_k$ matrices depending on $x^a$ 
and~$\h_i^{\a}$.

In~\cite{LP07} it was shown that all Ward-model field equations are satisfied 
if one takes
\begin{equation}\label{3.18}
S_k\=S_k(w_k, \h^i_k) \und
T_k\=\biggl\{\prod\limits_{l=1}^{k-1}\Bigl({\bf 1}_n\ -\ \frac{
\mu_{k-l}-\bar\mu_{k-l}}{\mu_{k-l}-\bar\mu_{k}}\ P_{k-l}\Bigr)\biggr\}\ S_k 
\end{equation} 
with
\begin{equation}\label{3.19}
w_k:= x + 
\sfrac12(\bar\mu_k{-}\bar\mu_k^{-1})\,y+\sfrac12(\bar\mu_k{+}\bar\mu_k^{-1})\,t
\quad\mbox{and}\quad
\h^i_k:=\h_i^{1} + \bar\mu_k\h_i^{2}
\quad\mbox{for}\quad k=1,\ldots,m\ .
\end{equation} 
Substituting (\ref{3.18}) into (\ref{3.15}), we obtain from (\ref{3.16})
the solution
\begin{equation}\label{3.20}
\Phi_m \= \j^{-1}_m(\zeta{=}0) \= 
\prod\limits_{k=1}^{m}\bigl({\bf 1}_n-\rho_kP_k\bigr)
\qquad\mbox{with}\quad\rho_k=1-\frac{\mu_k}{\bar\mu_k}\ .
\end{equation} 
Furthermore, from (\ref{3.13}) we read off
\begin{equation}\label{3.21}
\Acal =\sum\limits_{k=1}^m(\mu_{k}{-}\bar\mu_{k})\pa_xP_k\ ,\quad
\Bcal =\sum\limits_{k=1}^m(\mu_{k}{-}\bar\mu_{k})(\pa_t{-}\pa_y)P_k
\quad\mbox{and}\quad
\Ccal^i =\sum\limits_{k=1}^m(\mu_{k}{-}\bar\mu_{k})\pa_1^iP_k\ .
\end{equation} 
Thus, the solutions of the noncommutative $\Ncal$-extended integrable U($n$)
chiral model in 2+1 dimensions described by the simple-pole ansatz 
(\ref{3.16})--(\ref{3.18}) are parametrized by the set $\{S_k\}^m_1$ of 
matrix-valued functions of $w_k$ and $\h^i_k$ and by the pole positions $\m_k$.

\bigskip

\section{Configuration of two noncommutative plane waves}
\noindent
The solutions constructed in the previous section have solitonic character 
when all the functions $\{S_1\}^m_1$ are rational 
(see e.g.~\cite{Ward88, Io, TU, LP1, LP2, LP07}). 
For the dressing ansatz (\ref{3.16})  with pairwise distinct $\m_k$ 
it was shown that no scattering occurs in the $\Ncal$-extended $m$-soliton 
configuration~\cite{LP07}. However, it is known that in the bosonic 
commutative~\cite{Leese} and noncommutative~\cite{B, IU} cases 
the choice of exponentials for $\{S_1\}^m_1$ leads to
a configuration of $m$ {\it plane waves\/} which do feature interaction. 
It is natural to expect that $\Ncal$-extended plane-wave configurations 
have the same properties. We will demonstrate this on the example of a 
two-wave configuration which is a particular solution of the noncommutative 
$\Ncal{=}2$ supersymmetric U(2) Ward model. Note that the properties of 
a solution describing $m$~extended waves essentially depends on $\m_k\in\C$ 
and the parameters in $\{S_k\}^m_1$, and a complete study of the interaction 
of such waves is far from a trivial matter~\cite{Leese}.
That is why we restrict ourselves to a special form of $S_k$ and a choice of 
parameters which simplifies the analysis.

\medskip

\noindent
{\bf Extended wave solution.} \ 
Let us take $\m$ to be purely imaginary,
\begin{equation}\label{4.1}
\m =-\im p\qquad\mbox{with}\qquad p>1\ .
\end{equation}
Then from (\ref{3.19}) for $\Ncal{=}2$ we obtain
\begin{equation}\label{4.2}
w\=x\ +\ \sfrac{\im}{2p}\bigl((p^2{+}1)\,y\,+\,(p^2{-}1)\,t \bigr) \und
\h\=\h^1\,+\,\im\,p\,\h^2\ .
\end{equation}
We choose $T=S=S(w,\h)$ in the form
\begin{equation}\label{4.3}
T\=\biggl(\begin{matrix}1+\h\ve\\[4pt] \ep^{bw}\end{matrix}\biggr) \ ,
\end{equation}
where $\ve$ is a Grassmann-odd parameter and $b=b^x+\im\,b^y$ is a 
complex number. The form (\ref{4.3}) is obviously not the most general,
but it extends the simplest bosonic wave ansatz by a nilpotent term.

Our choice (\ref{4.1})--(\ref{4.3}) then yields
\begin{equation}\label{4.4}
P\=T\,(T^\+T)^{-1}T^\+\=\begin{pmatrix}
\frac{\ab\,\a}{\ab\,\a + \ep^{2{\tx} - {\tt}}}&
\frac{\a\, \ep^{\tx}\ep^{-i\ty}}{\ab\,\a + \ep^{2\tx }}\\[12pt]
\frac{\ab\, \ep^{\tx}\ep^{i\ty}}{\ab\,\a + \ep^{2\tx }}&
\frac{\ep^{2\tx +\tt}}{\ab\,\a + \ep^{2\tx + \tt}}
\end{pmatrix}
\end{equation}
with the abbreviations
\begin{equation}\label{4.5}
\tx\ :=\ b^xx\ -\ \sfrac{b^y}{2p}\bigl((p^2{+}1)\,y\,+\,(p^2{-}1)\,t\bigr) \und
\ty\ :=\ b^yx\ +\ \sfrac{b^x}{2p}\bigl((p^2{+}1)\,y\,+\,(p^2{-}1)\,t\bigr)\ ,
\end{equation}
\begin{equation}\label{4.6}
\tt\ :=\ \sfrac{|b|^2}{2p}(p^2{+}1)\,\th\ ,\qquad 
\a\ :=\ 1+\h\,\ve \und \quad\ab\=1+\bar\h\,\bar\ve\ .
\end{equation}
For the U(2)-valued superfield~$\Phi$ which is by construction a solution
to~(\ref{2.2}), we finally find the one-wave configuration
\begin{equation}\label{4.7}
\Phi \= {\bf 1}_n - 2P \=\begin{pmatrix}
\frac{\ep^{2{\tx} - {\tt}} - \ab\,\a}{ \ep^{2{\tx} - {\tt}}+\ab\,\a }&
-\frac{2\a\, \ep^{\tx -i\ty}}{\ep^{2\tx} +\ab\,\a }\\[12pt]
-\frac{2 \ab\, \ep^{\tx + i\ty}}{\ep^{2\tx }+\ab\,\a }&
-\frac{ \ep^{2\tx +\tt}-\ab\,\a }{\ep^{2\tx + \tt}+\ab\,\a }
\end{pmatrix}\ .
\end{equation}
Notice that all expressions in (\ref{4.4}) and (\ref{4.7}) are formed with
Moyal star multiplication.

The wave described by (\ref{4.3})--(\ref{4.7}) simply moves at constant 
velocity which can be shown by the same arguments as in~\cite{Leese}.
Moreover, the wave front lies along
\begin{equation}\label{4.8}
\tilde x =0 \ ,
\end{equation}
which for fixed time is a straight line in the $xy$-plane.
{}From (\ref{4.4}) one can see that (cf.~\cite{IU}) 
\begin{equation}\label{4.9}
\lim\limits_{\tx\to -\infty} P\= \begin{pmatrix}1&0\\0&0\end{pmatrix} \und
\lim\limits_{\tx\to +\infty} P\= \begin{pmatrix}0&0\\0&1\end{pmatrix}\ ,
\end{equation}
corresponding to the large-time limits $t\to \pm\infty$ for $b^y<0$ 
and finite $x,y$.\footnote{
For $b^y>0$ we simply have the correspondence 
$\tx\to\pm\infty\ \Leftrightarrow\ t\to\mp\infty$.} 
Thus, we get the asymptotics
\begin{equation}\label{4.10}
\Phi_{\pm\infty}\=\lim\limits_{t\to \pm\infty} \Phi \= 
\pm\begin{pmatrix}1&\!\!\phantom{-}0\\0&\!\!-1\end{pmatrix}\ ,
\end{equation}
which corresponds to straight wave moving far away from the $t{=}0$ line
\begin{equation}\label{4.11}
b^xx\ -\ \sfrac{b^y}{2p}(p^2{+}1)\,y\=0 
\end{equation}
on either side.

\medskip

\noindent
{\bf Interacting waves.} \ 
Now we consider two waves defined by formulae similar to 
(\ref{4.1})--(\ref{4.3}). Namely, we choose
\begin{equation}\label{4.12}
\m_k=-\im\,p_k \und \h_k\=\h^1+\im\,p_k\h^2
\qquad\mbox{with}\quad p_2>p_1>1\ ,
\end{equation}
\begin{equation}\label{4.13}
w_k\=x\ +\ \sfrac{\im}{2p_k}\bigl((p^2_k{+}1)\,y+(p^2_k{-}1)\,t\bigr) \und
\a_k=1+\h_k\ve_k\ ,
\end{equation}
\begin{equation}\label{4.14}
S_k\=\biggl(\begin{matrix}\a_k\\[4pt] \ep^{b_kw_k}\end{matrix}\biggr)
\qquad\mbox{for}\quad k=1,2\ ,
\end{equation}
where $\ve_k$ are Grassmann-odd parameters and $b_k=b_k^x+\im\,b_k^y$ 
are complex numbers. We also introduce
\begin{equation}\label{4.15}
\tx_k\ :=\ b^x_kx-\sfrac{b^y_k}{2p_k}\bigl((p_k^2{+}1)\,y+(p_k^2{-}1)\,t\bigr)
\quad\mbox{and}\quad
\ty_k\ :=\ b^y_kx+\sfrac{b^x_k}{2p_k}\bigl((p_k^2{+}1)\,y+(p_k^2{-}1)\,t\bigr)
\ ,
\end{equation}
so that
\begin{equation}\label{4.16}
b_kw_k \= \tx_k + \im\,\ty_k \und
[\tx_k, \ty_k]=\im\,\tt_k \qquad\mbox{with}\qquad
\tt_k\ :=\ \sfrac{|b_k|^2}{2p_k}(p_k^2{+}1)\,\th\ .
\end{equation}
The relations (\ref{3.18}) read
\begin{equation}\label{4.17}
T_1\=S_1 \und T_2\=\bigl({\bf 1}_2\ -\ \sfrac{2p_1}{p_1+p_2}P_1\bigr)\,S_2\ ,
\end{equation}
from which we construct the matrices
\begin{equation}\label{4.18}
P_k\=T_k\,(T_k^\+T_k)^{-1}T_k^\+\quad\mbox{for}\quad k=1,2 \und
\Phi \=({\bf 1}_2-2P_2)\,({\bf 1}_2-2P_1)\ ,
\end{equation}
arriving at a two-wave configuration.

Let us move with the second wave. This means that we consider points 
around its wave front defined by the equation
\begin{equation}\label{4.19}
\tx_2\ \equiv\ 
b^x_2 x\ -\ \sfrac{b^y_2}{2p_2}\bigl((p_2^2{+}1)\,y+(p_2^2{-}1)\,t\bigr)\=0\ ,
\end{equation}
which is a line in the $xy$-plane moving in time. For a proper choice of 
parameters keeping $\tx_2$ finite while $\tx_1\to\pm\infty$, asymptotically 
the first wave will be far away from the second wave on either side.
Specifically for $b_1^y<0$, (\ref{4.18}) and (\ref{4.9}) give us
\begin{equation}\label{4.20}
{T_2}|_{t,\tx_1\to-\infty}\ \to\ \left\{{\bf 1}_2 - \sfrac{2p_1}{p_1+p_2}
\begin{pmatrix}1&0\\0&0\end{pmatrix}\right\}
\biggl(\begin{matrix}\a_2\\[4pt] \ep^{b_2w_2}\end{matrix}\biggr) \=
\ep^\g\ \biggl(\begin{matrix}\a_2\\[4pt] \ep^{b_2w_2-\g}\end{matrix}\biggr)\ ,
\end{equation}
\begin{equation}\label{4.21}
{T_2}|_{t,\tx_1\to+\infty}\ \to\ \left\{{\bf 1}_2 - \sfrac{2p_1}{p_1+p_2}
\begin{pmatrix}0&0\\0&1\end{pmatrix}\right\}
\biggl(\begin{matrix}\a_2\\[4pt] \ep^{b_2w_2}\end{matrix}\biggr) \= 
\phantom{\ep^\g\ }
\biggl(\begin{matrix}\a_2\\[4pt] \ep^{b_2w_2+\g}\end{matrix}\biggr)\ ,
\end{equation}
where
\begin{equation}\label{4.22}
\ep^\g\ :=\ \frac{p_2-p_1}{p_2+p_1}\ .
\end{equation}
As a consequence, we arrive at
\begin{equation}\label{4.23}
\begin{aligned}
\Phi |_{t\to\pm\infty}\ \to\ &\pm
\begin{pmatrix}
\frac{\ep^{b_2w_2+\bb_2\wb_2\pm 2\g-\tt_2} - \ab_2\a_2}
{\ep^{b_2w_2+\bb_2\wb_2\pm 2\g-\tt_2}+\ab_2\a_2 }&
-\frac{2\a_2\ep^{\bb_2\wb_2\pm\g}}{\ep^{b_2w_2+\bb_2\wb_2\pm2\g}+\ab_2\a_2}
\\[12pt]
-\frac{2\ab_2\ep^{b_2w_2\pm\g}}{\ep^{b_2w_2+\bb_2\wb_2\pm2\g}+\ab_2\a_2}&
-\frac{\ep^{b_2w_2+\bb_2\wb_2\pm2\g+\tt_2}-\ab_2\a_2}
{\ep^{b_2w_2+\bb_2\wb_2\pm2\g+\tt_2}+\ab_2\a_2}
\end{pmatrix}\begin{pmatrix}1&\!\!\phantom{-}0\\0&\!\!-1\end{pmatrix} 
\\[12pt]
\= &\pm \begin{pmatrix}
\frac{\ep^{2\tx_2\pm2\g-\tt_2} - \ab_2\a_2}
{\ep^{2\tx_2\pm 2\g-\tt_2}+\ab_2\a_2}&
\frac{2\a_2\ep^{\tx_2-\im\ty_2\pm \g}}{\ep^{2\tx_2\pm 2\g} +\ab_2\a_2}
\\[12pt]
-\frac{2 \ab_2\ep^{\tx_2+\im\ty_2\pm \g}}{\ep^{2\tx_2\pm \g}+\ab_2\a_2}&
\frac{\ep^{2\tx_2\pm 2\g +\tt_2}-\ab_2\a_2}
{\ep^{2\tx_2\pm 2\g+\tt_2}+\ab_2\a_2}
\end{pmatrix}\ .
\end{aligned}
\end{equation}
Comparing the $t\to\pm\infty$ limits of this configuration, we see that 
(up to sign) $\Phi|_{t\to +\infty}$ deviates from $\Phi|_{t\to -\infty}$ 
only by the phase shift $\ b_2w_2{-}\g\mapsto b_2w_2{+}\g$. This coincides 
with the $\Ncal{=}0$ commutative result found in~\cite{Leese}. 
By symmetry, both waves experience the same phase shift of
\begin{equation}
2\g \= 2\,\ln\,\frac{p_2-p_1}{p_2+p_1}\ .
\end{equation}
Note that the explicit asymptotic form of $\Phi$ depends on~$\th$ 
but the phase shift does not. 
Although our two waves interact in a rather simple way, 
their dynamics depends essentially on the parameters $p_k$, $b_k$ and $\ve_k$. 
The $\Ncal{=}0$ waves are recovered by putting $\ve_k=0$, i.e.~$\a_k=1$. 
In the $\Ncal$-extended case the choice of $S_k$ may be more general 
than~(\ref{4.14}), leading to more involved interaction dynamics.

\bigskip

\section{Two interacting solitons}
\noindent
{\bf One-soliton configuration.} \ 
According to the general formalism discussed in
section 3, the time-dependent configuration for $m{=}1$ 
and $\mu$ not necessarily being purely imaginary simplifies to
\begin{equation}\label{5.1}
\j\= {\bf 1}_n + \frac{\m - \bar\m}{\zeta -\m}P \und
\Phi\=\j^{-1}(\zeta{=}0)\= {\bf 1}_n + \frac{\bar\m -\m}{\m}P
\end{equation}
with
\begin{equation}\label{5.2}
P\=T\,(T^\+T)^{-1}T^\+ \und T\=T(w_1,\h_1^i)\ ,
\end{equation}
where $w_1$ and  $\h_1^i$ are given in (\ref{3.19}). This configuration will 
describe a moving soliton if the $n{\times}r$ matrix $T$ depends on $w_1$
rationally (cf.~e.g.~\cite{Ward88, LP1, LP07}). 
For $\m=-\im$ we encounter the static case, where
\begin{equation}\label{5.3}
\j \= {\bf 1}_n - \frac{2\im}{\zeta +\im}P\ , \quad 
\Phi \= {\bf 1}_n - 2P\ ,\quad 
w_1\=z\equiv x+\im y \quad\mbox{and}\quad 
\h_1^i\=\h^i\equiv \h_i^1 + \im \h_i^2
\end{equation}
with $P$ and $T=T(w_1, \h_1^i)=T(z,\h^i)$ given in (\ref{5.2}). 
The $n{\times}n$ matrix superfields $\Acal$, $\Bcal$ and $\Ccal^i$ 
from the linear system (\ref{3.4}) and eqs.~(\ref{3.7}) are expressed 
in terms of $P$ as (cf.~(\ref{3.21}))
\begin{equation}\label{5.4}
\Acal \= -2\im\,\pa_x P\ ,\qquad \Bcal \= 2\im\,\pa_y P \und
\Ccal^i \= -2\im\,\pa_1^i P\ .
\end{equation}

We want to `dress' the static solution (\ref{5.3}) of the field 
equations~(\ref{2.2}) to produce a time-dependent interacting two-soliton
configuration of the Moyal-deformed supersymmetric Ward model. 
It is known in the non-supersymmetric case that soliton interactions
appear only when higher-order poles in~$\z$ are considered for
the dressing ansatz~(\ref{3.16})~\cite{Io, IZ, LP2}.
The simplest such situation occurs for a double pole at $\zeta=-\im$.
Therefore, we take the static configuration given by (\ref{5.3}) 
and~(\ref{5.4}) as our seed solution and consider the dressing transformation
\begin{equation}\label{5.5}
\j\ \mapsto\ \tilde\j \= ({\bf 1}_n-\sfrac{2\im}{\zeta+\im}\tilde P)\,\j \=
({\bf 1}_n-\sfrac{2\im}{\zeta+\im}\tilde P)\,
({\bf 1}_n-\sfrac{2\im}{\zeta+\im}P) \=
{\bf 1}_n -\sfrac{2\im}{\zeta+\im}(P{+}\tilde P) -
\sfrac{4}{(\zeta+\im)^2}\tilde P\,P\ .
\end{equation}
{}From the reality condition~(\ref{3.11}) we obtain 
the restrictions $\ \tilde P^2=\tilde P\ $ and $\ \tilde P^\+=\tilde P\ $, 
qualifying $\tilde P$ as a hermitian projector
\begin{equation}\label{5.6}
\tilde P\=\tilde T\,(\tilde T^\+\tilde T)^{-1}\tilde T^\+
\end{equation}
built from some $n{\times}\tilde r$ matrix~$\tilde T$. 
In the following we choose $r=1=\tilde r$,
i.e.~we consider rank one projectors $P$ and~$\tilde P$.

Demanding that $\tilde\j$ is again a solution of the linear equations 
(\ref{3.4}) with some new superfields $\tilde\Acal$, $\tilde\Bcal$ 
and $\tilde\Ccal^i$, we derive
\begin{equation}\label{5.7}
\begin{aligned}
&\tilde\Acal\=\tilde\j(\pa_t{+}\pa_y{-}\zeta\pa_x)\tilde\j^\+ \= 
({\bf 1}_n{-}\sfrac{2\im}{\zeta{+}\im}\tilde P)\Acal
({\bf 1}_n{+}\sfrac{2\im}{\zeta{-}\im}\tilde P)+
({\bf 1}_n{-}\sfrac{2\im}{\zeta{+}\im}\tilde P)(\pa_t{+}\pa_y{-}\zeta\pa_x)
({\bf 1}_n{+}\sfrac{2\im}{\zeta{-}\im}\tilde P)\ , \\[8pt]
&\tilde\Bcal \=\tilde\j(\pa_x{-}\zeta\pa_t{+}\zeta\pa_y)\tilde\j^\+ \= 
({\bf 1}_n{-}\sfrac{2\im}{\zeta{+}\im}\tilde P)\Bcal
({\bf 1}_n{+}\sfrac{2\im}{\zeta{-}\im}\tilde P){+}
({\bf 1}_n{-}\sfrac{2\im}{\zeta{+}\im}\tilde P)(\pa_x{-}\zeta\pa_t{+}\zeta\pa_y)
({\bf 1}_n{+}\sfrac{2\im}{\zeta{-}\im}\tilde P)\ , \\[8pt]
&\tilde\Ccal^i\=\tilde\j(\pa_2^i{-}\zeta\pa_1^i)\tilde\j^\+ \= 
({\bf 1}_n{-}\sfrac{2\im}{\zeta{+}\im}\tilde P)\Ccal^i
({\bf 1}_n{+}\sfrac{2\im}{\zeta{-}\im}\tilde P)+
({\bf 1}_n{-}\sfrac{2\im}{\zeta{+}\im}\tilde P)(\pa_2^i{-}\zeta\pa_1^i)
({\bf 1}_n{+}\sfrac{2\im}{\zeta{-}\im}\tilde P)\ .
\end{aligned}
\end{equation}
The poles at $\zeta=\pm\im$ on the right-hand side of these equations
have to be removable since $\tilde\Acal$, $\tilde\Bcal$ and $\tilde\Ccal^i$
are independent of $\zeta$. Putting to zero the corresponding residues, 
we find the conditions
\begin{subequations}
\begin{eqnarray}\label{5.10} 
&&({\bf 1}_n{-}\tilde P)\,\bigl(\pa_\zb\tilde T+(\pa_\zb P)\,\tilde T\bigr)\=0
\ , \\[4pt] \label{5.11}
&&({\bf 1}_n{-}\tilde P)\,\bigl(\pa_t\tilde T-2\im(\pa_z P)\,\tilde T\bigr)\=0
\ , \\[4pt] \label{5.12}
&&({\bf 1}_n{-}\tilde P)\,\bigl(\sfrac12(\pa_1^i{+}\im\pa_2^i)\tilde T +
(\pa_1^iP)\,\tilde T\bigr) \=0 \ .
\end{eqnarray}
\end{subequations}
After constructing a projector $\tilde P$ via a solution $\tilde T$
of these equations, we obtain a solution (\ref{5.5}) of the linear
equations (\ref{3.13}) and, hence, a new (dressed) superfield 
\begin{equation}\label{5.13}
\tilde\Phi \= \tilde\j^{-1}(\zeta{=}0) \= 
({\bf 1}_n - 2P)\,({\bf 1}_n - 2\tilde P)
\end{equation}
obeying the field equations~(\ref{2.2}).

\medskip

\noindent
{\bf Explicit nonabelian solution.} \ 
In order to generate an explicit example solving (\ref{5.10})--(\ref{5.12}), 
we specialize to the group U(2) (i.e.~choose $n{=}2$) and 
take as a one-soliton seed configuration
\begin{equation}\label{5.14}
P\=T\,(T^\+T)^{-1}T^\+ \qquad\mbox{with}\qquad 
T\=\biggl(\begin{matrix} 1 \\[4pt] f(z,\h^i) \end{matrix}\biggr)\ ,
\end{equation}
where implicit $\star$ products are still assumed everywhere. Inspired by the 
known form of~$\tilde T$ in the bosonic case~\cite{Io, LP2}, 
we make the ansatz
\begin{equation}\label{5.15}
\tilde T \= T\ +\ T_\bot (T_\bot^\+T_\bot )^{-1}g \qquad\mbox{with}\qquad 
T_\bot\=\biggl(\begin{matrix}\overline{f(z,\h^i)}\\[4pt]-1\end{matrix}\biggr)
\end{equation}
being orthogonal to $T$, i.e.
\begin{equation}\label{5.17}
T^\+T_\bot \=0\qquad\Rightarrow\qquad P\,T_\bot\=0 \und
{\bf 1}_2 - P \= T_\bot (T_\bot^\+T_\bot^{} )^{-1}T_\bot^\+\ ,
\end{equation}
and with $g(t,z,\zb,\h^i,\bar\h^i)$ being a superfield to be determined.

Substituting (\ref{5.15}) into (\ref{5.10}), we get
\begin{equation}\label{5.18}
\pa_{\zb}g\=0\qquad\Rightarrow\qquad g\=g(t,z,\h^i,\bar\h^i)\ . 
\end{equation}
{}From (\ref{5.11}) it follows that
\begin{equation}\label{5.19}
\pa_{t}g\=-2\im\,\pa_zf\qquad\Rightarrow\qquad 
g\=-2\im \bigl(t\,\pa_zf + H(z,\h^i,\bar\h^i)\bigr)\ .
\end{equation}
Finally, from (\ref{5.12}) we obtain
\begin{equation}\label{5.20}
\bar\pa_i g \= -\pa_i f \qquad\Rightarrow\qquad
g\=-2\im \bigl(t\,\pa_zf + h(z,\h^i)\bigr)\ +\ \bar\h^i \pa_i f \ ,
\end{equation}
where $h(z,\h^i)$ is an arbitrary function of $z$ and $\h^i$,
and we have used the abbreviations (cf.~(\ref{5.3}))
\begin{equation}
\pa_i\ :=\ \sfrac12(\pa_1^i - \im\pa_2^i) \und 
\bar\pa_i\ :=\ \sfrac12(\pa_1^i + \im\pa_2^i) \ .
\end{equation}

For further analysis we expand $f$ and $h$ in $\h^i$,
\begin{equation}\label{5.21}
f(z,\h^i)\=f_0(z)+\h^if_i(z)+\ldots \und h(z,\h^i)\=h_0(z)+\h^ih_i(z)+\ldots\ .
\end{equation}
If we restrict ourselves to a bosonic subsector, studied in~\cite{Io, LP2}, 
then the choice
\begin{equation}\label{5.22} 
f_0\=z \und h_0\=z^2
\end{equation}
yields a configuration of two lumps centered at $z=\pm\sqrt{-t}$, which
for negative times accelerate symmetrically along the $x$-axis towards 
the origin $z{=}0$ of the Moyal plane, interact at small~$t$, and 
decelerate to infinity along the $y$-axis for positive times. 
Thus, a head-on collision of these lumps results in a $90^\circ$~scattering. 
For the general superfield solution given by (\ref{5.13})--(\ref{5.15})
and~(\ref{5.20}), the analysis seems much more complicated even when 
$\Ncal{=}2$. However, for any $\Ncal{\le}8$ and $f_0$, $h_0$ chosen as above, 
the bosonic core of the solution behaves in the above-described way. 
Hence, two $\Ncal$-extended lumps carrying fermionic degrees of freedom
can interact in the Moyal plane.
We postpone a full-fledged scattering analysis of these supersymmetric 
configurations to future work.

\medskip

\noindent
{\bf Explicit abelian solution.} \
A genuinely novel feature of noncommutative sigma models is the appearance
of {\it abelian\/} solitons, i.e.~nontrivial solutions for the group U(1).
To describe these, one must employ the Moyal-Weyl correspondence and
represent the noncommutativity by {\it operator-valued\/} functions of only 
$(t|\h^\a_i,\th^{i\a})$ instead of $\C$-number functions of 
$(t,z,\bar z|\h^\a_i,\th^{i\a})$ subject to $\star$~multiplication.
These operators act on an auxiliary Fock space~$\Fcal$ spanned by the basis
\begin{equation}
|\ell\> \= \sfrac{(a^\+)^\ell}{\sqrt{\ell!}}\,|0\> \qquad\textrm{for}\qquad
\ell=0,1,2,\ldots \und a\,|0\> \= 0 \qquad\textrm{where}\qquad [a\,,a^\+]\=1\ .
\end{equation}
Here, $\sqrt{2\th}\,a$ is the operator corresponding to the coordinate
function~$z$, and likewise for the hermitian conjugate. In this setting, 
projectors of finite rank~$r$ in the total space~$\C^n\otimes\Fcal$ 
decompose as
\begin{equation}
P \= |T\>\,\<T|T\>^{-1}\<T| \ ,\qquad\textrm{where}\quad
|T\> \= \bigl( |T_1\>, |T_2\>, \ldots, |T_r\> \bigr)
\end{equation}
denotes a row of $r$~kets from $\C^n\otimes\Fcal$. 
In the following we take~$n=1$ and $r=1$ (and drop the index).

It was demonstrated in~\cite{LP07} that the U(1)~solutions are based on
a coherent state 
\begin{equation}
|T\> \= |\a(\h^i)\> \= \ep^{\a(\h^i)a^\+} |0\>
\end{equation}
with a Grassmann-valued parameter~$\a$. We may always translate a static 
soliton to the origin of the Moyal plane, which amounts to dropping the
body part of~$\a$. Considering $\Ncal{=}2$ supersymmetry, i.e.~a single
complex Grassmann-odd coordinate~$\h$, this implies
\begin{equation}
\a(\h) \= \h\,\eps
\end{equation}
with a Grassmann-odd parameter~$\eps$, and thus
\begin{equation} \label{Tket}
|T\> \= |\h\eps\> \= |0\>\ +\ \h\eps\,|1\>\ .
\end{equation}
The corresponding projector is easily computed, and the final static abelian
rank-one one-soliton configuration reads (in $\star$-product formulation)
\begin{equation}
\Phi \= 1\ -\ 4\,\ep^{-|z|^2/\th}\,\Bigl\{ 
1\ +\ \sfrac{2\bar z}{\sqrt{2\th}}\,\h\eps\ +\ 
\sfrac{2z}{\sqrt{2\th}}\,\bar\eps\bar\h\ +\ 
2\bigl( \sfrac{|z|^2}{\th}{-}1\bigr)\,\eps\bar\eps\h\bar\h \Bigr\}\ .
\end{equation}

To construct a time-dependent abelian two-soliton solution~(\ref{5.13}),
we must dress the seed solution based on~(\ref{Tket}) with a factor of
$1{-}2\tilde P$ based on a second ket~$|\tilde T\>$. Again taking the
rank~$\tilde r{=}1$, we solve~(\ref{5.10})--(\ref{5.12}) and find
\begin{equation}
|\tilde T\>\=|T\>\ +\ |T^1_\bot\>\,g_1(t)\ +\ |T^2_\bot\>\,g_2(t)
\qquad\textrm{with}\qquad \<T\,|\,T^{1,2}_\bot\>\=0 \ ,
\end{equation}
where $\ g_1(t)=1\ $ and $\ g_2(t)=-\im t\,\sqrt{2/\th}\ $ multiply the kets
\begin{equation}
|T^1_\bot\> \= (\h\bar\eps + \eps\bar\h)\,|1\> \und
|T^2_\bot\> \= \bar\h\bar\eps\,|0\>\ +\ (1-\eps\bar\eps\h\bar\h)\,|1\>\ 
            +\ \sqrt{2}\,\h\eps\,|2\> \ ,
\end{equation}
respectively.
As in the bosonic case, the time dependence drops out for $t\to\pm\infty$,
and the two limits yield the same asymptotic configuration, which is supported 
near the origin. Hence, this configuration describes a two-soliton bound state
dressed with a fermionic degree of freedom.

\bigskip

\noindent
{\bf Acknowledgements}

\medskip

\noindent
The authors are grateful to A.D.~Popov for fruitful discussions and useful 
comments. O.L. thanks P.~Aschieri, P.~Kulish and J.~Madore for discussions.
T.A.I.~acknowledges the Heisenberg-Landau program and the Russian Foundation
for Basic Research (grant 06-01-00627-a) for partial support and the Institut 
f\"ur Theoretische Physik der Leibniz Universit\"at Hannover for its 
hospitality. The work of O.L.~is partially supported by the Deutsche 
Forschungsgemeinschaft.
\bigskip
%\newpage


\begin{thebibliography}{99}
%\addtolength{\itemsep}{-4pt}

\bibitem{DH}
  M.R.~Douglas and C.M.~Hull,
  %``D-branes and the noncommutative torus,''
  J. High Energy Phys. {\bf 02} (1998) 008
  [hep-th/9711165].
  %%CITATION = HEP-TH 9711165;%%

\bibitem{Chu}
  C.S.~Chu and P.M.~Ho,
  %``Noncommutative open string and D-brane,''
  Nucl. Phys. B {\bf 550} (1999) 151
  [hep-th/9812219].
  %%CITATION = HEP-TH 9812219;%%
\bibitem{Sch}
  V.~Schomerus,
  %``D-branes and deformation quantization,''
  J. High Energy Phys. {\bf 06} (1999) 030
  [hep-th/9903205].
  %%CITATION = HEP-TH 9903205;%%

\bibitem{SW}
  N.~Seiberg and E.~Witten,
  %``String theory and noncommutative geometry,''
  J. High Energy Phys. {\bf 09} (1999) 032
  [hep-th/9908142].
  %%CITATION = HEP-TH 9908142;%%

\bibitem{LPS1}
  O.~Lechtenfeld, A.D.~Popov and B.~Spendig,
  %``Open N = 2 strings in a B-field background and noncommutative self-dual
  %Yang-Mills,''
  Phys.\ Lett.\ B {\bf 507} (2001) 317 
  [hep-th/0012200].
  %%CITATION = HEP-TH 0012200;%%

\bibitem{LPS2}
  O.~Lechtenfeld, A.D.~Popov and B.~Spendig, \\{}
  %``Noncommutative solitons in open N = 2 string theory,''
  J. High Energy Phys. {\bf 06} (2001) 011
  [hep-th/0103196].
  %%CITATION = HEP-TH 0103196;%%

\bibitem{Ward88}
  R.S.~Ward,
  %``Soliton solutions in an integrable chiral model in 2 + 1 dimensions,''
  J. Math. Phys. {\bf 29} (1988) 386;
  %``Classical solutions of the chiral model, unitons, and holomorphic vector
  %bundles,''CITATION = JMAPA,29,386;%%
  Commun. Math. Phys.  {\bf 128} (1990) 319.

\bibitem{Io}
  R.S.~Ward,
  %``Nontrivial scattering of localized solitons in a (2+1)-dimensional
  %integrable system,''
  Phys.\ Lett.\  A {\bf 208} (1995) 203;\\
  %%CITATION = PHLTA,A208,203;%%
  T.A.~Ioannidou,
  % ``Soliton Solutions and Nontrivial Scattering in an Integrable Chiral Model
  %in (2+1) Dimensions,''
  J.\ Math.\ Phys.\  {\bf 37} (1996) 3422 
  [hep-th/9604126].
  %%CITATION = JMAPA,37,3422;%%

\bibitem{IZ}
  T.A.~Ioannidou and W.J.~Zakrzewski,
  %``Solutions of the modified chiral model in (2+1) dimensions,''
  J.\ Math.\ Phys.\  {\bf 39}  (1998) 2693
  [hep-th/9802122].
  %%CITATION = JMAPA,39,2693;%%

\bibitem{Leese}
  R.~Leese,
  %``EXTENDED WAVE SOLUTIONS IN AN INTEGRABLE CHIRAL MODEL IN
  %(2+1)-DIMENSIONS,''
  J.\ Math.\ Phys.\  {\bf 30} (1989) 2072.
  %%CITATION = JMAPA,30,2072;%%

\bibitem{Manton}
  T.~Ioannidou and N.S.~Manton,
  %``The energy of scattering solitons in the Ward model,''
  Proc.\ Roy.\ Soc.\ Lond.\  A {\bf 461} (2005) 1965
  [hep-th/0409168];\\
  %%CITATION = PRSLA,A461,1965;%%
  M.~Dunajski and N.S.~Manton,
  %``Reduced dynamics of Ward solitons,''
  Nonlinearity {\bf 18} (2005) 1677
  [hep-th/0411068].
  %%CITATION = NOLIN,18,1677;%%

\bibitem{TU}
  B.~Dai and  C.L. Terng, \\
  ``Backlund transformations, Ward solitons, and unitons", math.DG/0405363;\\
  B.~Dai, C.L.~Terng and K.~Uhlenbeck, \\
  ``On the space-time monopole equation", math.DG/0602607.

\bibitem{LP1}
  O.~Lechtenfeld and A.D.~Popov,
  %``Noncommutative multi-solitons in 2+1 dimensions,''
  J. High Energy Phys. {\bf 11} (2001) 040
  [hep-th/0106213].
  %%CITATION = HEP-TH 0106213;%%

\bibitem{LP2}
  O.~Lechtenfeld and A.D.~Popov,
  %``Scattering of noncommutative solitons in 2+1 dimensions,''
  Phys.\ Lett.\ B {\bf 523} (2001) 178
  [hep-th/0108118].
  %%CITATION = HEP-TH 0108118;%%
\bibitem{B}
  S.~Bieling,
  %``Interaction of noncommutative plane waves in 2+1 dimensions,''
  J.\ Phys.\ A {\bf 35} (2002) 6281
  [hep-th/0203269].
  %%CITATION = HEP-TH 0203269;%%
\bibitem{W1}
  M.~Wolf,
  %``Soliton antisoliton scattering configurations in a noncommutative sigma
  %model in 2+1 dimensions,''
  J. High Energy Phys. {\bf 06} (2002) 055
  [hep-th/0204185].
  %%CITATION = HEP-TH 0204185;%%
\bibitem{IU}
  M.~Ihl and S.~Uhlmann,
  %``Noncommutative extended waves and soliton-like configurations in N = 2
  %string theory,''
  Int. J. Mod. Phys. A {\bf 18} (2003) 4889
  [hep-th/0211263].
  %%CITATION = HEP-TH 0211263;%%

\bibitem{Klawunn}
  M.~Klawunn, O.~Lechtenfeld and S.~Petersen,\\
  %``Moduli-space dynamics of noncommutative Abelian sigma-model solitons,''
  J. High Energy Phys. {\bf 06} (2006) 028
  [hep-th/0604219].
  %%CITATION = JHEPA,0606,028;%%

\bibitem{Penati}
  O.~Lechtenfeld, L.~Mazzanti, S.~Penati, A.D.~Popov and L.~Tamassia,\\
  %``Integrable noncommutative sine-Gordon model,''
  Nucl. Phys. B {\bf 705} (2005) 477
  [hep-th/0406065].
  %%CITATION = HEP-TH 0406065;%%

\bibitem{Ward}
  R.S.~Ward,
  %``Integrable and solvable systems, and relations among them,''
  Phil.\ Trans.\ Roy.\ Soc.\ Lond.\ A {\bf 315} (1985) 451;
  %%CITATION = PTRSA,A315,451;%%
  ``Multidimensional integrable systems,''
  %\href{http://www.slac.stanford.edu/spires/find/hep/www?irn=1793730}{SPIRES entry}
  In: {\it Field Theory, Quantum Gravity and Strings}, Eds. H.J. De Vega,
  N. Sanchez, Vol. 2, p.106, 1986.

\bibitem{Mason}
  L.J.~Mason and G.A.J.~Sparling,
  %Nonlinear Schr\ödinger and Korteweg-De Vries are reductions of 
  %self-dual Yang-Mills,
  Phys.\ Lett.\ A {\bf 137} (1989) 29;
  %%CITATION = PHLTA,A137,29;%%
  %Twistor correspondences for the soliton hierarchies,
  J.\ Geom.\ Phys.\  {\bf 8} (1992) 243;
  %%CITATION = JGPHE,8,243;%%
  L.J.~Mason and M.A.~Singer,
  %The twistor theory of equations of KdV type. 1,
  Commun.\ Math.\ Phys.\  {\bf 166} (1994) 191.
  %%CITATION = CMPHA,166,191;%%

\bibitem{Chakravarty}
  S.~Chakravarty, M.J.~Ablowitz and P.A.~Clarkson,
  %``Reductions of self-dual Yang-Mills fields and classical systems,''
  Phys.\ Rev.\ Lett.\  {\bf 65} (1990) 1085;
  %%CITATION = PRLTA,65,1085;%%
  M.J.~Ablowitz, S.~Chakravarty and L.A.~Takhtajan,
  %A self-dual Yang-Mills hierarchy and its reductions to integrable 
  %systems in 1+1 dimensions and 2+1 dimensions,
  Commun.\ Math.\ Phys.\  {\bf 158} (1993) 289;
  %%CITATION = CMPHA,158,289;%% 
  S.~Chakravarty, S.L.~Kent and E.T.~Newman,
  %``Some reductions of the self-dual Yang-Mills equations to integrable systems
  %in (2+1)-dimensions,''
  J.\ Math.\ Phys.\  {\bf 36} (1995) 763;\\
  %%CITATION = JMAPA,36,763;%%
  M.J.~Ablowitz, S.~Chakravarty and R.G.~Halburd,
  %``Integrable systems and reductions of the self-dual Yang-Mills equations,''
  J.\ Math.\ Phys.\  {\bf 44} (2003) 3147. 
  %%CITATION = JMAPA,44,3147;%%

\bibitem{Ivanova:1992tk}
  T.A.~Ivanova and A.D.~Popov,
  %``Soliton equations and self-dual gauge fields,''
  Phys.\ Lett.\ A {\bf 170} (1992) 293;\\
  %%CITATION = PHLTA,A170,293;%%
  %``Some new integrable equations from the self-dual Yang-Mills equations,''
  Phys.\ Lett.\ A {\bf 205} (1995) 158
  [hep-th/9508129];
  %``Self-dual Yang-Mills fields in $D=4$ and integrable systems in $1\le D\le
  %3$,''
  Theor.\ Math.\ Phys.\  {\bf 102} (1995) 280;\\
  %%CITATION = TMPHA,102,280;%%
  %%CITATION = HEP-TH 9508129;%%
  M.~Legar\'e and A.~D.~Popov,
  %``Reductions of a Lax pair for self-duality equations of the Yang-Mills
  %model,''
  JETP Lett.\  {\bf 59} (1994) 883;
  %%CITATION = JTPLA,59,883;%%
  %``Lax pairs of integrable equations in 1 <= D <= 3 dimensions as reductions
  %of the Lax pair for the self-dual Yang-Mills equations,''
  Phys.\ Lett.\ A {\bf 198} (1995) 195.
  %%CITATION = PHLTA,A198,195;%%

\bibitem{HamToda}
  M.~Hamanaka and K.~Toda,
  %``Towards noncommutative integrable systems,''
  Phys.\ Lett.\ A {\bf 316} (2003) 77
  [hep-th/0211148];\\
  %%CITATION = HEP-TH 0211148;%%
  %``Noncommutative Burgers equation,''
  J.\ Phys.\ A {\bf 36} (2003) 11981
  [hep-th/0301213];\\
  %%CITATION = HEP-TH 0301213;%%
  M.~Hamanaka,
  %``On reductions of noncommutative anti-self-dual Yang-Mills equations,''
  Phys.\ Lett.\ B {\bf 625} (2005) 324
  [hep-th/0507112];  
  %%CITATION = HEP-TH 0507112;%%

\bibitem{Dimakis}
  A.~Dimakis and F.~Mueller-Hoissen,
  %``Bicomplexes and integrable models,''
  J.\ Phys.\ A {\bf 33} (2000) 6579 
  [nlin.si/0006029];\\
  M.T.~Grisaru, L.~Mazzanti, S.~Penati and L.~Tamassia,\\
  %``Some properties of the integrable noncommutative sine-Gordon system,''
  J. High Energy Phys. {\bf 04} (2004) 057
  [hep-th/0310214];\\ 
  %M.T.~Grisaru, L.~Mazzanti, S.~Penati and L.~Tamassia,
  %``Some properties of the integrable noncommutative sine-Gordon system,''
  J. High Energy Phys. {\bf 04} (2004) 057
  [hep-th/0310214];\\
  %%CITATION = HEP-TH 0310214;%%
  I.~Cabrera-Carnero,
  %``Abelian Toda field theories on the noncommutative plane,''
  J. High Energy Phys. {\bf 10} (2005) 071
  [hep-th/0503147].
  %%CITATION = HEP-TH 0503147;%%

\bibitem{Wit}
  E.~Witten,
  %``Perturbative gauge theory as a string theory in twistor space,''
  Commun.\ Math.\ Phys.\  {\bf 252} (2004) 189
  [hep-th/0312171].
  %%CITATION = HEP-TH 0312171;%%

\bibitem{Popov}
  A.D.~Popov and C.~Saemann,
  %``On supertwistors, the Penrose-Ward transform and $Ncal$=4 super Yang-Mills
  %theory,''
  Adv.\ Theor.\ Math.\ Phys.\  {\bf 9} (2005) 931
  [hep-th/0405123];\\
  %%CITATION = HEP-TH 0405123;%%
  O.~Lechtenfeld and C.~Saemann,
  %``Matrix models and D-branes in twistor string theory,''
  J. High Energy Phys. {\bf 03} (2006) 002
  [hep-th/0511130];\\
  %%CITATION = HEP-TH 0511130;%%
  A.D.~Popov and M.~Wolf,
  %``Hidden symmetries and integrable hierarchy of the $\Ncal{=}4$ 
  %  supersymmetric  Yang-Mills equations,'' 
  Commun.\ Math.\ Phys.\ {\bf 275} (2007) 685 
  [hep-th/0608225].
  %%CITATION = CMPHA,275,685;%%

\bibitem{Bedford}
  J.~Bedford, C.~Papageorgakis and K.~Zoubos,\\
  ``Twistor Strings with Flavour,''
  arXiv:0708.1248 [hep-th].
  %%CITATION = ARXIV:0708.1248;%%

\bibitem{sigma8}
  A.D.~Popov,
  Phys.\ Lett.\ B {\bf 647} (2007) 509 [hep-th/0702106].
  %%CITATION = HEP-TH/0702106;%%

\bibitem{LP07}
  O.~Lechtenfeld and A.D.~Popov,
  J. High Energy Phys. {\bf 06} (2007) 065 [arXiv:0704.0530 [hep-th]]
  %%CITATION = JHEPA,0706,065;%%

\bibitem{Ohta}
  M.~Hamanaka, Y.~Imaizumi and N.~Ohta,
  %``Moduli space and scattering of D0-branes in noncommutative super
  %Yang-Mills theory,''
  Phys.\ Lett.\ B {\bf 529} (2002) 163 [hep-th/0112050].

\bibitem{Z}
  V.E.~Zakharov and A.V.~Mikhailov,
  %``Relativistically invariant two-dimensional models in field theory
  %integrable by the inverse problem technique. (In Russian),''
  Sov.\ Phys.\ JETP {\bf 47} (1978) 1017;\\
  %%CITATION = SPHJA,47,1017;%%
  V.E.~Zakharov and A.B.~Shabat,
  %``Integration of nonlinear equations by the inverse scattering method.
  Funct.\ Anal.\ Appl.\  {\bf 13} (1979) 166;\\
  %%CITATION = FAAPB,13,166;%%
  P.~Forg\'acs, Z.~Horv\'ath and L.~Palla,
  %``Solution generating technique for self-dual monopoles,''
  Nucl.\ Phys.\ B {\bf 229} (1983) 77;\\
  %%CITATION = NUPHA,B229,77;%%
  K.~Uhlenbeck,  J. Diff. Geom. {\bf 30} (1989) 1;\\
  %%CITATION = JDGEA,30,1;%%  
  O.~Babelon and D.~Bernard,
  %``Dressing symmetries,''
  Commun.\ Math.\ Phys.\  {\bf 149} (1992) 279
  [hep-th/9111036].
  %%CITATION = HEP-TH 9111036;%%

\bibitem{reviews}
  A.~Konechny and A.S.~Schwarz,
  %``Introduction to M(atrix) theory and noncommutative geometry,''
  Phys. Rept.  {\bf 360} (2002) 353
  [hep-th/0012145, hep-th/0107251];
  %%CITATION = HEP-TH 0012145;%%
  %%CITATION = HEP-TH 0107251;%%
  M.R.~Douglas and N.A.~Nekrasov,
  %``Noncommutative field theory,''
  Rev. Mod. Phys.  {\bf 73} (2001) 977
  [hep-th/0106048];\\
  %%CITATION = HEP-TH 0106048;%%
  R.J.~Szabo,
  %``Quantum field theory on noncommutative spaces,''
  Phys. Rept.  {\bf 378} (2003) 207
  [hep-th/0109162].
  %%CITATION = HEP-TH 0109162;%%

\end{thebibliography}
\end{document}